\documentclass{kluwer}    

\usepackage[dvips]{graphicx}

\newdisplay{guess}{Conjecture}

\def\gee{ \, \lower 1mm\hbox{$\,{\buildrel > \over{\scriptstyle\scriptstyle\sim} }\displaystyle \,$}}
\def\lee{ \, \lower 1mm\hbox{$\,{\buildrel < \over{\scriptstyle\scriptstyle\sim} }\displaystyle \,$}}
\def\|{\partial}

\def\varkappa {{\scriptstyle\partial}\! e}

\begin{document}
\begin{article}

\begin{opening}
\title{Disk-to-halo mass ratio evaluations based on the numerical models of collisionless disks}
\author{Zasov \surname{Anatoly V.}\email{zasov@sai.msu.ru}}
\author{Khoperskov \surname{Alexander V.}\email{khopersk@sai.msu.ru}}
\author{Tyurina \surname{Nataly V.}\email{tiurina@sai.msu.ru}}
 \runningauthor{A.V. Zasov,  A.V. Khoperskov, N.V. Tyurina} \runningtitle{Disk-to-halo mass ratio}
\institute{Moskow State University, Russia}
\date{April 20, 2004}

\begin{abstract}

We propose that the lower bound of the stellar radial
velocity dispersion $c_r$ of an equilibrium stellar
disk is determined by the gravitational stability
condition. We compared the estimates of stellar
velocity dispersion at radii $r>(1.5 - 2)\cdot R_0$
(where $R_0$ is the photometric radial scalelength of
a disk), found in the literature, with the minimal
values of $c_r$  necessary  for the disk to be in a
stable state, using the results of numerical
simulations of 3D collisional disks. This approach
enables to estimate an upper limit of the local
surface density and (if $R_0$ is known) a total masses
of a disk and a dark halo. We argue that the old
stellar disks of spiral galaxies with active star
formation usually have the velocity dispersions which
are close to the expected marginal values. A rough
values of disk-to-total mass ratios (within the fixed
radius) are found for about twenty spiral galaxies.
Unlike spirals, the disks of the ``red'' Sa--S0
galaxies are evidently ``overheated'': their radial
dispersion of velocities at $r \simeq 2 R_0$ exceeds
significantly the marginal values for gravitational
stability.
\end{abstract}

\end{opening}
\section{Introduction}
It is usually accepted that substantial amounts of dark matter are needed to explain flat rotation curves in the outer
regions of disk galaxies. However the fraction of a dark matter belonging to a disk and to a halo is still an open
question. Several different methods were proposed how to get a separate estimate of a mass or a local density of a
disk. One may try, for example,
\begin{itemize}
\item  To decompose  a rotation curve of a galaxy into
spherical and disk --- related components; \item To
assume that the mass-to-luminosity ${M_d}/{L}$ ratio
of stellar disk is known from the models of stellar
population and to convert $L$ into $M_d$; \item  To
model either the formation or the kinematic properties
of such structural features as a bar or spiral arms;
\item To compare the observed scaleheight of gaseous
layer with the expected one for different $M_d$,
assuming that the velocity dispersion of gas is known;
\item  To use the available information about the
velocity dispersion of an old stellar disk population
(in addition to the rotation curve).
\end{itemize}

The first two approaches suffer from the well known ambiguities and usually cannot get the unique solution. The third
method needs the development of some theoretical background which will allow to connect the observational data with
theoretical expectations. The fourth method may be applied (with some inevitable assumptions) to edge-on galaxies only.
The applicability and the effectivety of the last method is still an open question. In this work we tried to verify to
what extent the observed velocity dispersion of an old disk population may be used to estimate the  local density or
total mass of a disk.

Galactic disks, consisting mostly of old stars, may be
considered as collisionless systems in
quasi-stationary equilibrium with a very slow
evolution (if to exclude the cases of a strong
interactions or mergings with neighbor systems). The
disk in equilibrium is characterized by certain radial
distribution of stellar velocity dispersions in the
plane of a disk $(c_r, c_\varphi)$, that ensures its
stability to gravitational perturbations. If the
stability criteria were known, it would allow to
develop a self-consistent model for the disk of real
galaxy  and to estimate its density when both the
rotational velocities and stellar velocity dispersions
are measured. It is convenient to describe a local
radial velocity dispersion $c_r$ in terms of
dimensionless Toomre parameter (which equals to unit
for the uniformly rotating thin disk, marginally
stable to radial perturbations):

 \begin{equation}\label{Formula-1}
 Q_T  =  c_r  \cdot
\varkappa / 3.36 G \sigma \,,
\end{equation}

 where
$\varkappa$ is the epicyclic frequency, and $\sigma$ is a non-perturbed local density of a disk. If the local value of
$c_r$ is known from observations, one may estimate the  local density from (1). To follow this way, one may either a)
to assume that Toomre stability parameter $Q_T$ is known from some theoretical stability criterium (see Zasov, 1985,
Bottema, 1993 and more recent papers, cited in Khoperskov et al., 2003), or to apply N-body models to galaxies to
verify that the observed velocity dispersion satisfies the condition of stable equilibrium (e.g. Zasov, 1985,
Khoperskov et al., 2001, Zasov et al., 2002). This method is rather cumbersome, but it may be used in a more simple
way if to find a minimum value of $Q_T$ of marginally stable disks which may be  applied to models with a wide range
of parameters.

\section{THE CHOICE OF   $Q_T$ FOR MARGINALLY STABLE DISKS.}

To find the minimum radial velocity dispersion sufficient to ensure stability of the disk against perturbations of
arbitrary shapes is highly important if, as some authors have suggested, real galaxies may be in a state of threshold
(marginal) stability. Note that in the general case, the old stellar population of a galactic disk can have an excess
velocity dispersion in the presence of other factors heating the disk (which may act both from inside and from
outside), which are not related directly to the gravitational instability.
 However, even in this case, the conditions for
marginal stability still provide a valuable information by yielding an upper limit for the mass of the disk that
enables it to be stable.

There are two different approaches used to  find the
threshold value of velocity dispersion:  either to
seek for analytical solution of the problem or to use
the numerical models to reproduce the observed
properties of real galaxies assuming different mass
distribution in the disk, bulge and halo. Together
with certain advantages over numerical simulations
(the mathematical rigorousness of the solutions in the
framework of the problem formulated), the analytical
approach to the dynamics of perturbations in a disk
and the conditions for its stability has   the
drawback that it can be implemented only for very
simple models (usually  2D disks and the local
perturbations are considered based on the analysis of
dispersion equation in the epicyclical approximation)
and can yield only coarse estimates for the parameters
of the disk when applied to real objects. Numerical
simulations of collisionless systems are more flexible
in terms of the choice of model. They make it possible
to go beyond simple two-dimensional models and
directly follow the development of perturbations in a
disk that is initially in equilibrium. However, this
approach has drawbacks of its own.
 The most serious problems of $N$-body simulations
include (1) certain inevitable mathematical
simplifications, and (2) the dependence of the final
state of the system (after it reaches
quasi-equilibrium) on the initial parameters, which
are poorly known for real galaxies.

 Both numerical and
analytical estimates lead to conclusion that the
marginal stability of a disk corresponds to $Q_T > 1$,
rising to $Q_T \ge $3 in the outer parts, although
different approaches may give different values for the
same disk parameters. We refer to our paper
(Khoperskov et al., 2003) for the more detailed
discussion.
 We analyzed there the conditions for the gravitational
stability of a three-dimensional collisionless disk
with exponential density profile, embedding in the
gravitational field of two rigid spherical components
--- a bulge and a halo, whose central concentrations
and radial scales were varied over wide ranges. The
initially weakly unstable disks in our models started
their evolution from the subcritical equilibrium
state. The results of dynamical simulations allowed to
determine the disk parameters at the stability limit
(when the velocity dispersion ceases to change, and
the disk reaches a quasi-steady state after 5 - 20
rotations of the outer edge of the disk). The
stability of the solutions against the choice of
computational method was verified by comparing results
for several models obtained by two very different
methods of computing the gravitational force: the
TREE-code method and direct ``particle-to-particle''
(PP) integration, in which each particle interacts
with each of the others, for $N = (20 - 80) \times
10^3$. A comparison of the two results revealed no
significant differences between the final disk states.

 We constructed different
numerical models (the number $N$ of equal mass particles was up to $500 \times 10^3$), where the ratios of halo-to
disk masses inside of a disk radius varied between 0.5 and 3. In the case of a low-mass or nonexistent halo, the
evolution of the disk is determined by the bar mode, and the disk is heated due to the formation of an
non-axisymmetric bar and associated two-armed spiral. Models with sufficiently compact bulge or massive halos do not
show any enhancement of the bar mode, however they develop a complex transient system of small-scale spiral waves. The
decrease of the amplitudes of these waves with time is accompanied by a transfer of rotational kinetic energy to the
chaotic component of the velocity, resulting in heating of the disk. In turn, the increase of the radial-velocity
dispersion $c_r$ slows down with decreasing wave amplitude. The heating virtually ceases after the decay of the
transient spiral waves. If the disk is initially cool ($Q_T \le 1 $),  its heating is  very efficient, and its
dynamical evolution clearly demonstrates that the wave-decay process has a certain inertia: the velocity dispersion is
already high enough to maintain the stability of the disk, however the spiral waves have not yet decayed (as is
confirmed by Fourier analysis of the density perturbations in the disk) and continue to heat the disk. Therefore, to
obtain the minimum velocity dispersion required for disk stability, we used an iterative algorithm, seeking for a
subcritical starting point to make the initial velocity dispersion approach the stability limit.

The other essential initial parameter is the disk
thickness or vertical-to-radial velocity dispersion
ratio. The thicker is the disk initially, the lower is
the minimum radial velocity dispersion $c_r$, which
determines its stability. This circumstance also shows
that the minimum critical dispersions in both
coordinates $z$ and $r$ are reached if the disk begins
to evolve from a subcritical state for both radial and
bending perturbations, slowly increasing radial and
vertical velocity dispersion at the initial state of
evolution. As expected, the minimum radial-velocity
dispersion at the end of the simulations (expressed in
units of the circular velocity) is higher in the
models where the relative mass of the halo is lower,
the initial disk thickness is less, and the degree of
differential disk rotation is higher.

\begin{figure}[!t]
{\includegraphics[width=0.495\hsize]{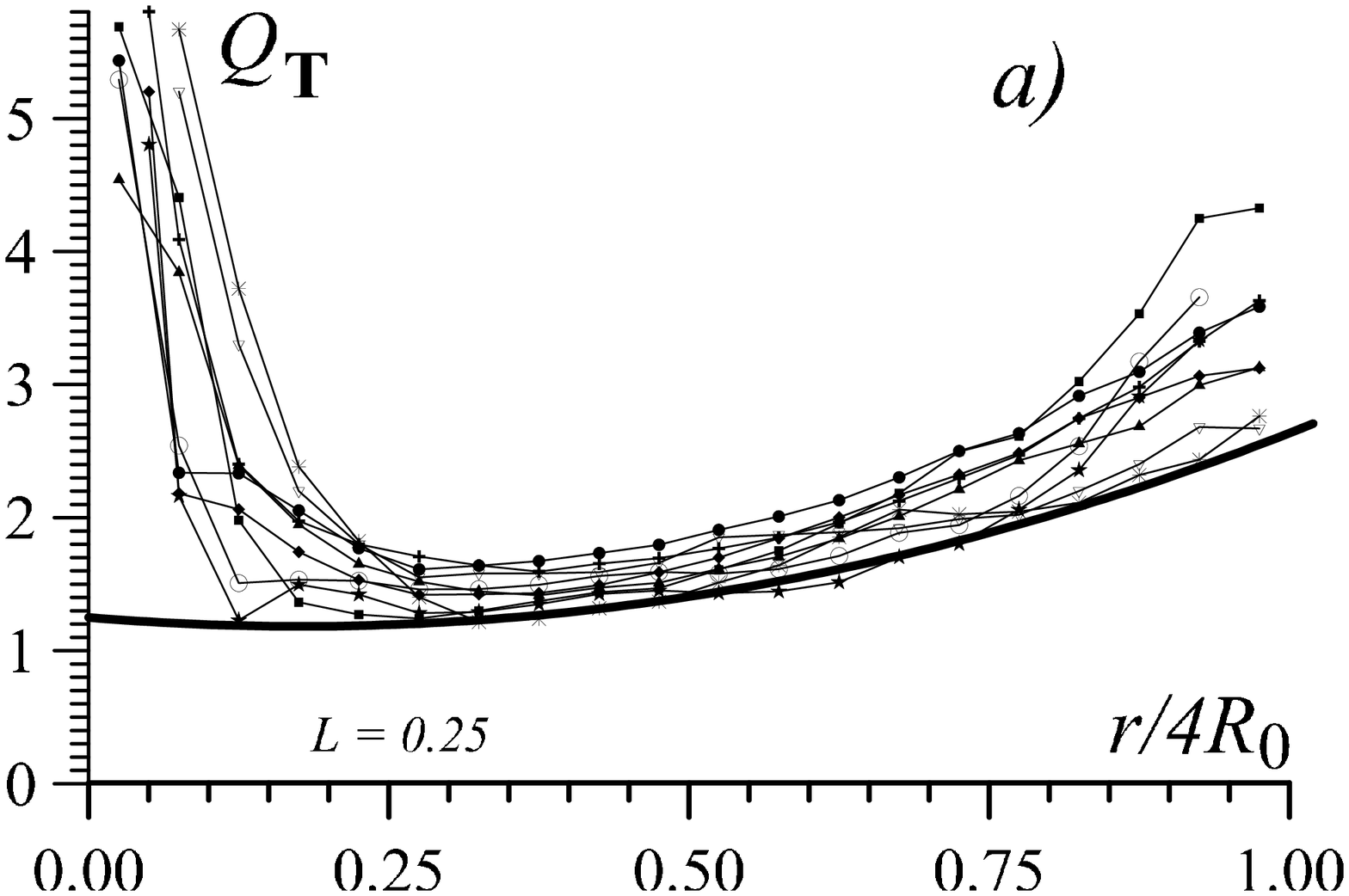}}
 \vskip -0.332\hsize \hskip 0.505\hsize
{\includegraphics[width=0.495\hsize]{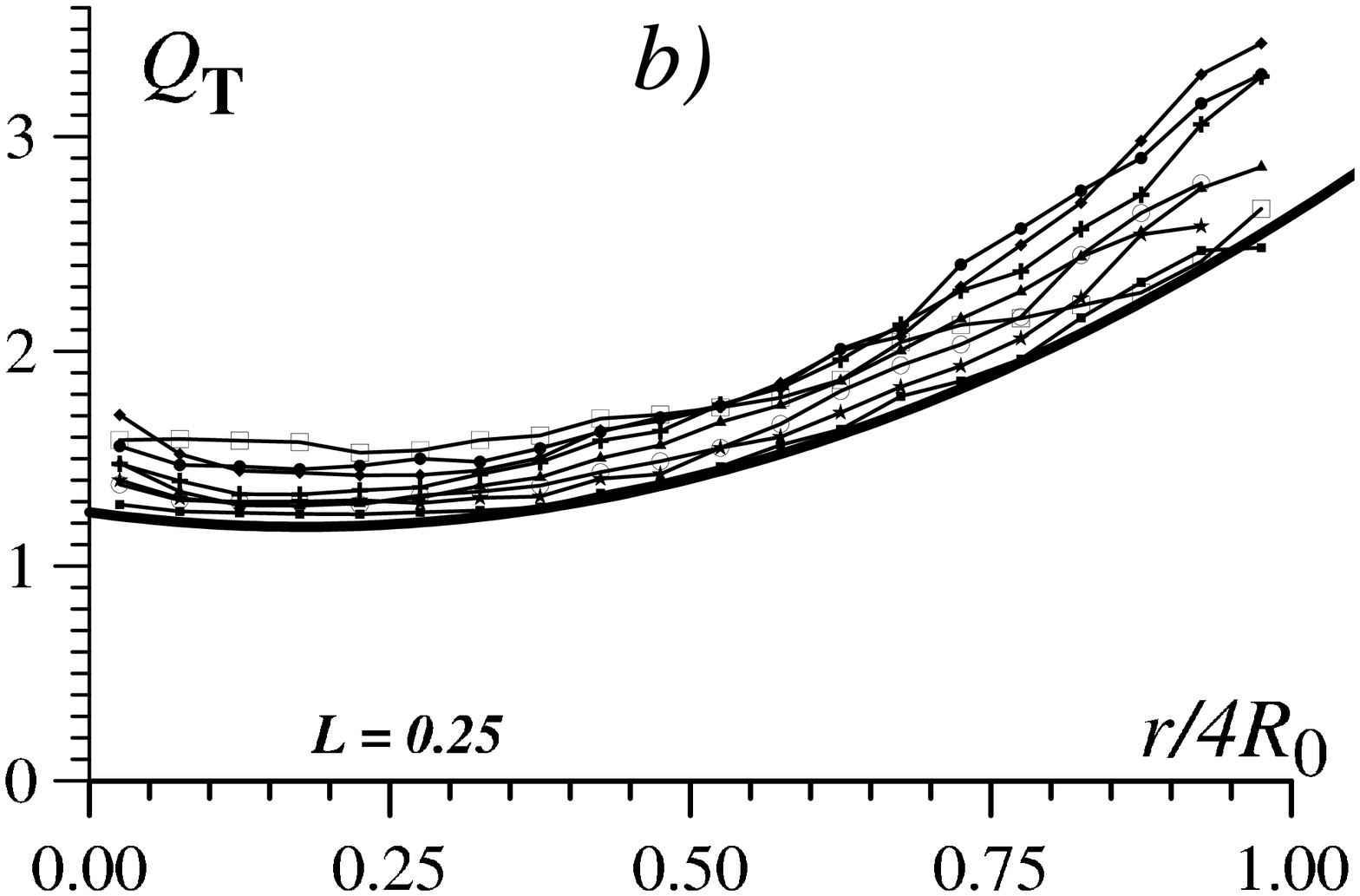}}
  \caption {
 Parameter Toomre $Q_T$ for marginally stable disks for different
 models with  bulges ({\it a}) and without bulges ({\it b}). Radius of disk
 corresponds to $r=4\,R_0$. } \label{PSAD-grav-inst-Fig1}
\end{figure}

The radial dependences of $Q_T$, calculated for our models, are different, being determined primarily by the relative
mass and degree of concentration of the spherical components. Yet  it is essential that in all cases we considered the
run of $Q_T(r)$ along the radius passes through a minimum $Q_T \simeq 1.2 - 1.6 $ just beyond the region controlled by
a bulge, at a galactocentric distance of $(1 - 2)\cdot R_0 $, where $R_0$ is the radial scale of a disk (see Figs 1{\it
a} and {\it b}), and this behavior depends only slightly on the choice of model. If a bulge is of low mass absent the
low value of $Q_T$ keeps down to the very centre (Fig. 1 b) This property can be used for a rough estimate of the
density (and, consequently, the mass) of a galactic disk (or to put limits on these quantities) from the observed
$c_r$ at these radial distances without the use of numerical simulations or analytical stability criteria.

\section{THE APPLICATION OF THE METHOD}
The measurements of stellar velocity dispersion
obtained by absorption line spectroscopy are usually
restricted by the central parts of galaxies due to
steep brightness gradient of stellar disks and the
difficuties of analysing the low intensity absorption
line spectra.  There are not so many galaxies where
the line-of-sight dispersion is measured at $r\ge
2R_0$. At this distance the brightness of a bulge is
usually negligible or at least is not overwhelming
(some exceptions may exist however among the early
type disky galaxies {\it Sa--S0}). For our purpose we
took the data obtained for spiral and lenticular
galaxies from the papers listed below:

\begin{itemize}
\item Shapiro et al (2003): NGC~1068, NGC~2460,
NGC~2775, NGC~4030;

 \item  Bershady et al (2002):
NGC~3982; \item Bottema (1993): NGC~1566, NGC~2613,
NGC~3198, NGC~5247, NGC~6340, NGC~6503;

 \item
Heraudeau et al. (1999): IC~750;

 \item Beltran et al.,
(2001): NGC~470, NGC~4419, NGC~7782;

 \item Simien,
Prugniel (2000, 2002): NGC~2962, NGC~3630, NGC~4143,
NGC~4203, NGC~4578, NGC~5273;

 \item
 Neistein et al. (1999):
NGC~584, NGC~2549, NGC~2768, NGC~3489, NGC~4251,
NGC~4649; NGC~4753, NGC~5866.
\end{itemize}

\begin{figure}[!t]
\centerline{\includegraphics[width=0.65\hsize]{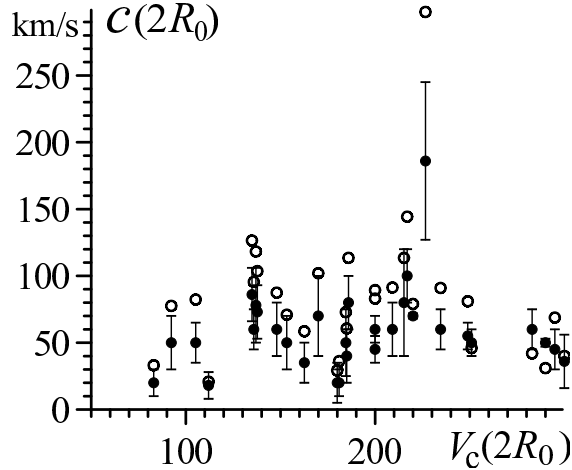}}
   \caption {
 A comparison of velocity dispersion at $r\simeq 2R_0$ with the
  velocity of rotation of galaxy.
 Filled circles are for the observed velocity dispersion $c_{obs}$,
 open circles --- for the estimates of radial component $c_r$.
 Error bars are given for  $c_{obs}$ only. } \label{PSAD-grav-inst-Fig2}
\end{figure}

The radial scalelengths $R_0$ (reduced to $H_0$ = 75 km/s/Mpc, if necessary) were taken from the cited papers or, if
they are absent there, from Baggett, 1998 or Grosbol, 1985. For our Galaxy, which  was added to the list above, we came
from $c_r = $ 38 km/s in the solar vicinity (Dehnen, Binney, 1998), at  radius $r_\odot \simeq 2.7 R_0$. In most cases
the estimations of the line-of-sight velocity dispersion $c_{obs}$ are related to the major axis of a galaxy. The
radial velocity dispersion $c_r$ was calculated from $c_{obs}$ using the equation
\begin{equation}\label2
c_r = c_{obs} \cdot \left[(c_\varphi / c_r)^2 \sin^2 i
+ (c_z/c_r)^2 \cos^2 i\right]^{-1/2} \,.
\end{equation}
 It was accepted that $c_\varphi/c_r =\varkappa/
2\Omega $ (epicyclic approximation), and $c_z/c_r
\simeq 1/2$. To get $\varkappa $ and $\Omega$ we used
the rotation curves obtained from gas emission lines.
These curves were taken either from the original
papers (see the papers cited above) or from the
references presented in the Catalogue of kinematically
resolved data (see HYPERLEDA database). If only the
``stellar'' rotation curve is available, we applied to
it the approximate correction for the asymptotic drift
at $r = 2 R_0$ (in the original paper by Neistein et
al., 2002 the corrected velocities are already given).

Fig.~2 reproduces the estimates of the observed velocity dispersion at $r \simeq 2R_0$ (filled symbols) and the
corresponding values of $c_r$ (open symbols) in comparison with the maximal velocity of rotation of galaxies. The
correlation between the velocity of rotation and the velocity dispersion is practically absent. Error bars are given
only for the observed values. These bars are rather of illustrative nature, being taken by eye from $c_{obs}(r)$
diagrams given in the original papers.
 They
demonstrate  rather low accuracy of the estimates, especially for  galaxies with low velocity dispersion, for which
the errors of  $c_{obs}$ and, hence the local density estimates $\sigma (2R_0)$, in some  cases may exceed a factor of
two. It means that the results we obtain from these data for individual galaxies may be considered only as rather
preliminary ones. Nevertheless it makes a sense to compare them with those expected for the marginally stable disks.

\begin{figure}[!t]
\centerline{\includegraphics[width=0.58\hsize]{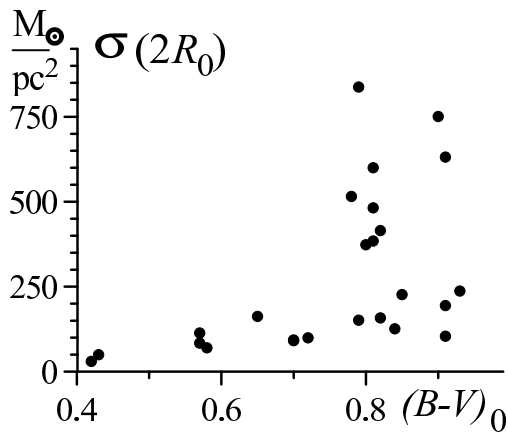}}
  \caption {
Threshold values of local surface density of galactic
disks at $r \simeq 2R_0$, corresponding to the
observed velocity dispersion, over the corrected total
color indices of galaxies. }
\label{PSAD-grav-inst-Fig3}
\end{figure}

In Fig.~3 we show the local surface densities of the
marginally stable disks of chosen galaxies at $r = 2
R_0$ , calculated for the adopted parameter $Q_T=$1.5.
They are plotted against the corrected values of color
indices of galaxies, taken from HYPERLEDA database. It
is worth to remember that if the disks are overstable,
that is if the velocity dispersion of stars exceeds
the threshold value for the gravitational stability,
these estimates may be considered as an upper bound of
the density at a given radius. The candidates to
galaxies with the overheated disks are several ``red''
galaxies (most of them are lenticulars) which stand
out in Fig.~3 by their incredibly high marginal
surface densities exceeding 400 $M_\odot/$pc$^2$,
whereas most of the other galaxies have $\sigma (2R_0)
\simeq (50-200)\, M_\odot/$pc$^2$. For comparison, in
solar neighborhood the column density of the galactic
disk does not exceed 60 $M_\odot / $pc$^2$
(Khoperskov, Tyurina, 2002).

 A similar conclusion
about the disk ``overheating'' of some galaxies follows from the estimates of the ratio of disk masses to total masses
of galaxies, which we describe as the indicative mass inside of the sphere with radius $r = 4R_0 $:
 \begin{equation}\label{Gr-Md-Mtot}
   \frac{M_d}{M_{t}} = \frac{\int\limits_0^{4 R_0} 2\pi r \sigma (r)\, dr}
   { 4 V^2 R_0/G} \,.
\end{equation}

  The galaxies of the ``red group'' mentioned above have an
unphysical estimates of the ratio $M_d / M_t> 1$, which means that their radial velocity dispersions definitely exceed
the marginal values even if to admit that the whole  mass of a galaxy contains in a disk. Hence either such galaxies
have very ``overstable'' stellar disks, or the measurements of dispersion were influenced by the light of the bulge,
which caused the overestimations of $c_{obs}$. Both versions are possible and in principle the situation has to be
analyzed separately for every single galaxy. It is worth noting however that for three of the presumably overheated
galaxies  (NGC~4251, NGC~4578 and NGC~5273) the existing estimates of  the velocity dispersion extend to $r
> 2R_0$, reaching the distances $r / R_0$ = 3.6, 3.3 and 3.1
correspondingly, that is their dispersion is obtained
at radii where the influence of a bulge is much lower
than at $r = 2R_0$. Although the uncertainty of $Q_T$
becomes more severe there, we may admit that its value
still does not exceed 3, as numerical models
demonstrate, which allows to obtain the upper limit of
the disk mass using    $c_{obs}$ at large radial
distances. However even in this case  the ratio $M_d /
M_t$ remains unphysically high  ($M_d  / M_t$ =1.3,
2.0 and 1.2 correspondingly for the galaxies in
question), that is the conclusion about the
``overheating'' of their disks is confirmed.

\begin{figure}[!t]
\centerline{\includegraphics[width=0.7\hsize]{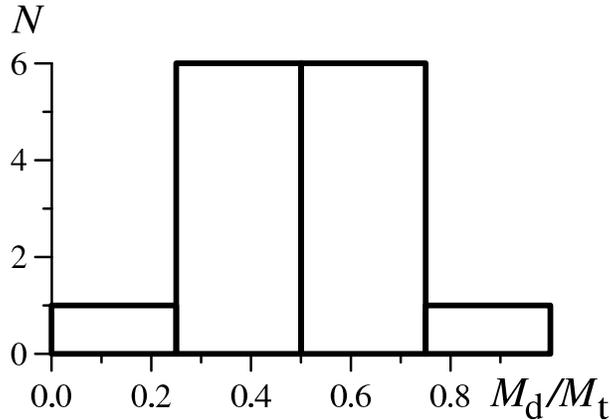}}
  \caption {
 A histogram of  $M_d$ over $M_t$ ratio, where $M_d$ is the mass of a
 marginally stable disk,  $M_t$ is the indicative total mass of a
 galaxy within the radius $r=4R_0$. } \label{PSAD-grav-inst-Fig4}
\end{figure}

It's essential that for all galaxies but two early type lenticulars (NGC~2768 and NGC~4203), which do not belong to a
group of red galaxies with a high marginal disk density at Fig.~3, the ratio $M_d / M_t$ < 1 (Fig.~4). It  means that
their radial velocity dispersions at $r = 2 R_0$, obtained from the observational data, cannot be explained without the
presence of massive haloes -- independently on whether the disks are marginally stable or overheated. Thence a dark
matter, if exists in a galaxy, cannot be concentrated in a disk. A fraction of mass of dark matter in galaxies seems to
change significantly from one galaxy to another.

 However we should have in mind that $M_d$   is no more than the crude
estimate of the upper limit of the disk mass.
 It is worth trying to verify whether the real disk mass is close to $M_d$.  If this proposition is correct, one can expect that the mass-to luminosity ratio for a disk is lower in galaxies with the less evolved stellar population. These galaxies should   possess lower color indices due to the presence of young stars.

\begin{figure}[!t]
\centerline{\includegraphics[width=0.6\hsize]{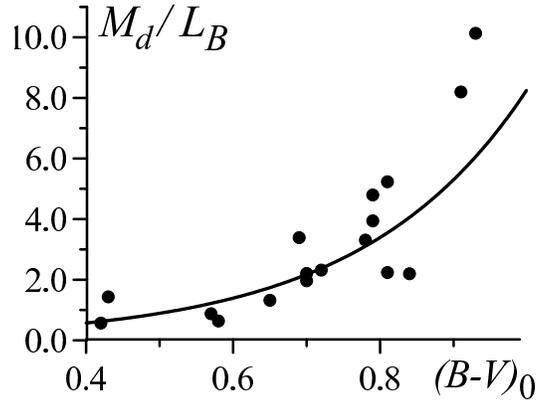}}
  \caption {
A relationship between mass-to-luminosity ratio
$M_d/L_B$ and the corrected color index for spiral
galaxies.  The  curve describes the  relationship
found from the evolution modeling of stellar system
(from Bell, deJong, 2001, closed box model).  }
\label{PSAD-grav-inst-Fig5}
\end{figure}

 In Fig.~5 we compare  the ratio $M_d/L_B$ with the corrected color
 index ($B-V$) for spiral galaxies of our sample, where the disk gives the main input to a total luminosity  (lenticular galaxies were omitted).  As the evolutionary models of stellar population show, the ratio $M_d/L_B$
increases along the sequence of color indices, weakly depending (for a given color) on the history of star formation
(Bell, de Jong, 2001). A curve drawn in Fig.~5 is model relationship, taken (without any normalization) from Bell, de
Jong 2001, which was obtained for a modified Salpeter Initial Mass Function and presented in an analytical form (see
their Table 3 for a ``closed box'' model). Two reddest galaxies are IC 750 (Sab) and NGC 2962 (S0a), the other galaxies
belong to later types. In spite of the significant spread of points (which is not surprising due to the crudeness of
the estimates), a general agreement between the model and the observed dependencies is evident:  masses of the disks
estimated under proposition of their marginal stability are close to those expected for galaxies with the observed
luminosity and colour. In enables to propose that in most spiral galaxies a radial velocity dispersions of old stars
in the disks  (at least for $r\simeq 2R_0$) is really close to the minimum values necessary for the disks to be
stable. If this is the case, one may conclude that the mechanisms of dynamical ``heating'' were not too efficient for
the late-type galaxies we considered after they reached a stable state.

This conclusion  evidently cannot be applied to the overstable disks of early type galaxies. It is possible that high
color indices (low star formation rate) and a low gas content, typical for these galaxies are caused by the same
events as the dynamical heating of their stellar disks (like a merging or a capture of small galaxies followed by the
fast gas consumption).

\section {CONCLUSION}
\begin{itemize}
\item Numerical modeling of 3D stellar disks evolving
from the sub-critical (to gravitational instability)
to stationary state allows to find the minimal value
of local parameter Toomre $Q_T \simeq 1.2 - 1.6$ which
reaches at the radial distance $r \simeq (1 - 2)\cdot
R_0$ for a wide ranges of masses and radial scales of
bulges, disks and halos of model galaxies.

\item The observed stellar velocity dispersion in the
disks of spiral galaxies well agrees with the
proposition that the radial velocity dispersion (at
least at $r \simeq 2R_0$) is close to the expected one
for the marginally stable disks.

\item The observed velocity dispersion in the disks of most of spiral galaxies we considered  makes the presence of a dark halo unavoidable for the disks to be gravitationally stable. Dark matter in a galaxiy, if exists, cannot be entirely concentrated in a disk.

\item Galaxies with the high color index ($B-V$) (most
of them are lenticular galaxies) may possess strongly
``overstable'' disks, with the radial velocity
dispersion exceeding  the threshold level for
gravitational instability. Hence the latter cannot be
responsible for the observed velocity dispersion.
\end{itemize}

The detailed discussion may be found in Zasov et al., AstL, 2004 (to be published).

\acknowledgements This work is supported by the
Russian Foundation for Basic Research through the
grants RFBR 04-02-16518 and by the Technology Program
``Research and Development in Priority Fields of
Science and Technology'' (contract 40.022.1.1.1101).

\end{article}


\begin{thebibliography}{}

\bibitem[Bell \& de Jong 1986]{Bell-Jong-2001}
 Bell E., de Jong R.S., 2001, Astrophys. J. \textbf{550}, 212

\bibitem[Baggett \& Baggett 1998]{Baggett-Baggett-1998}
 Baggett W.E., Baggett S.M., Anderson K.S.J., 1998, Astron. J., \textbf{116}, 1626

\bibitem[Beltran et al. 2001]{Beltran-etal-2001}
 Beltran J.C.V., Pizzella A.,
 Corsini E.M. et al., 2001, Astron. Astrophys. \textbf{374}, 391

\bibitem[Bershady et al. 2002]{Bershady-etal-2002}
 Bershady M., Verheijen M.,
 Anders D.A., 2002, ASP Conf. Proc. \textbf{275}, 43

\bibitem[Bottema 1993]{Bottema-1993}
 Bottema R., 1993, Astron. Astrophys. \textbf{275}, 16

\bibitem[Dehnen \& Binney 1998]{Dehnen-Binney-1998}
 Dehnen W., Binney J.J., 1998, MNRAS \textbf{298}, 387

\bibitem[Grosbol 1985]{Grosbol-1985}
 Grosbol P. J., 1985, Astron. Astrophys. Suppl. Ser. \textbf{60}, 261

\bibitem[Heraudeau 1999]{Heraudeau-etal-1999}
 Heraudeau Ph., Simien F., Maubon G., Prugniel Ph.,
 1999, Astron. Astrophys. Suppl. \textbf{136}, 509

\bibitem[]{}
 HYPERLEDA ~~http://www-obs.univ-lyon1.fr/hypercat/

\bibitem[Khoperskov et al. 2003]{Khoperskov-etal-2003}
 Khoperskov, A.V., Zasov, A.V., Tyurina, N.V. 2003, ARep, 47, 357

\bibitem[Khoperskov \& Tyurina 2003]{Khoperskov-Tyurina-2003}
 Khoperskov, A.V., Tyurina, N.V. 2003, ARep, 47, 443

\bibitem[Khoperskov et al. 2001]{Khoperskov-etal-2001}
 Khoperskov, A.V., Zasov, A.V., Tyurina, N.V. 2001, ARep, 45, 180

\bibitem[]{}
 Neistein E., Maoz D., Rix H,-W., Tonry J.L., 1999,
 Astron.J. \textbf{117}, 2666

\bibitem[]{}
 Shapiro K.L.. Gerssen J., van der Marel R.P., 2003,
 Astron. J., \textbf{126}, 2707

\bibitem[]{}
 Simien F., Prugniel Ph., 2000, Astron. Astrophys. Suppl. Ser. \textbf{145}, 263

\bibitem[]{}
 Simien F., Prugniel Ph., 2002, Astron. Astrophys. \textbf{384}, 371

\bibitem[]{}
 Zasov A.V., 1985, PAZh,  11, 730   

\bibitem[]{}
 Zasov, A.V., Bizyaev, D.V., Makarov, D.I., Tyurina, N.V. 2002, AstL, 28, 527

\end{thebibliography}
\end{document}